\pdfoutput=1
\documentclass[fleqn,usenatbib]{mnras}

\usepackage{newtxtext,newtxmath}

\usepackage[T1]{fontenc}

\DeclareRobustCommand{\VAN}[3]{#2}
\let\VANthebibliography\thebibliography
\def\thebibliography{\DeclareRobustCommand{\VAN}[3]{##3}\VANthebibliography}

\usepackage{graphicx}	
\usepackage{amsmath}	

\title[Jet in Mrk~590]{A parsec-scale faint jet in the nearby changing-look Seyfert galaxy Mrk~590}

\author[J. Yang et al.]{Jun Yang,$^{1}$\thanks{E-mail: jun.yang@chalmers.se}, 
Ilse van Bemmel,$^{2}$
Zsolt Paragi,$^{2}$
S. Komossa,$^{3}$
Feng Yuan,$^{4}$
Xiaolong Yang,$^{4}$
Tao An,$^{4}$
\and
J.~Y. Koay,$^{5}$
C. Reynolds,$^{6}$
J.~B.~R. Oonk,$^{7,8,9}$ 
Xiang Liu$^{10}$
and Qingwen Wu$^{11}$
\\
$^{1}$Department of Space, Earth and Environment, Chalmers University of Technology, Onsala Space Observatory, SE-439 92 Onsala, Sweden \\
$^{2}$Joint Institute for VLBI ERIC (JIVE), Postbus 2, NL-7990 AA Dwingeloo, the Netherlands \\
$^{3}$Max-Planck-Insitut f\"ur Radioastronomie, Auf dem H\"ugel 69, 53121 Bonn, Germany \\
$^{4}$Shanghai Astronomical Observatory, Key Laboratory of Radio Astronomy, Chinese Academy of Sciences, 200030 Shanghai, China \\
$^{5}$Institute of Astronomy and Astrophysics, Academia Sinica, Section 4, Roosevelt Rd., Taipei 10617, Taiwan \\
$^{6}$CSIRO Astronomy and Space Science, Kensington 6151, Australia \\
$^{7}$SURFsara, PO Box 94613, NL-1090 GP Amsterdam, the Netherlands \\
$^{8}$Leiden Observatory, Leiden University, PO Box 9513, NL-2300 RA Leiden, the Netherlands \\
$^{9}$Netherlands Institute for Radio Astronomy (ASTRON), NL-7991 PD Dwingeloo, the Netherlands \\
$^{10}$Xinjiang Astronomical Observatory, Key Laboratory of Radio Astronomy, Chinese Academy of Sciences,150 Science 1-Street, 830011 Urumqi, China \\
$^{11}$Department of Astronomy, School of Physics, Huazhong University of Science and Technology, Wuhan 430074, China \\
}
\date{Accepted XXX. Received YYY; in original form ZZZ}

\pubyear{2015}

\begin{document}
\label{firstpage}
\pagerange{\pageref{firstpage}--\pageref{lastpage}}
\maketitle

\begin{abstract}
Broad Balmer emission lines in active galactic nuclei (AGN) may display dramatic changes in amplitude, even disappearance and re-appearance in some sources. As a nearby galaxy at a redshift of $z = 0.0264$, Mrk~590 suffered such a cycle of Seyfert type changes between 2006 and 2017. Over the last fifty years, Mrk~590 also underwent a powerful continuum outburst and a slow fading from X-rays to radio wavelengths with a peak bolometric luminosity reaching about ten per cent of the Eddington luminosity. To track its past accretion and ejection activity, we performed very long baseline interferometry (VLBI) observations with the European VLBI Network (EVN) at 1.6~GHz in 2015. The EVN observations reveal a faint ($\sim$1.7~mJy) radio jet extending up to $\sim$2.8~mas (projected scale $\sim$1.4~pc) toward north, and probably resulting from the very intensive AGN activity. To date, such a parsec-scale jet is rarely seen in the known changing-look AGN. The finding of the faint jet provides further strong support for variable accretion as the origin of the type changes in Mrk~590. 

\end{abstract}




\section{Introduction}
\label{sec:intro}
Variations of the mass-accretion rate may give rise to different classes of emission-line nuclei of galaxies \citep[e.g.][]{Elitzur2014, Noda2018}. This scenario has been further supported by the findings of some optical changing-look active galactic nuclei (AGN) that cannot be explained as a consequence of variable absorption, e.g. the Seyfert galaxies Mrk~590 \citep{Denney2014} and Mrk~1018 \citep{McElroy2016, Noda2018}, and the quasar SDSS~J015957.64$+$003310.5 \citep{LaMassa2015}. Low optical linear polarisations \citep[13 changing-look quasars,][]{Hutsemkers2019} and large variations in the mid-infrared luminosity \citep[10 changing-look AGN,][]{Sheng2017} are inconsistent with the scenario of the dust obscuration. Recently, a few large samples of changing-look AGN have been selected \citep[e.g.][]{Runco2016, MacLeod2016, Yang2018}. If these dramatic type changes are mainly due to variations of the ionising continuum luminosity, changing-look AGN would play a key role in probing the complex accretion-ejection activity \citep[e.g.][]{Yuan2014, Noda2018} of super massive black holes (SMBHs).    

The existing observations show that intensive accretion events may trigger episodic ejections and launch (mildly) relativistic jets at speeds $\ga$0.1~$c$ \citep[e.g.][]{Marscher2002, Fender2009}. The accretion-ejection activity has been observed in the outbursts of many Galactic stellar-mass black hole X-ray binaries in particular during the transition from the X-ray low to high states \citep[e.g.][]{Fender2009, Yang2010, Yang2011, Miller-Jones2019}, a few tidal disruption events of SMBHs \citep[e.g.][]{Yang2016, Mattila2018}, some nearby AGN \citep[e.g.][]{Marscher2002, Argo2015} and short-lived radio sources \citep[5--20 yr, ][]{Mooley2016, Woowska2017}. VLBI observations of a few SMBHs accreting close to or above the Eddington accretion rate also reveal significant pc-scale jet activities while with faint radio cores \citep[e.g.][]{Yang2019, Yang2020, Yang2021}. Some theoretical models have been proposed to explain episodic ejections \citep[e.g.][]{Wu2009, Yuan2009}. 

Mrk~590 (alternatively, NGC~863) is a face-on Seyfert spiral galaxy at a redshift of $z = 0.0264$. It was traditionally classified as a Seyfert 1 galaxy \citep{Osterbrock1993} while recently found to undergo remarkable type changes \citep{Denney2014, Mathur2018, Raimundo2019}. Its broad optical H$\beta$ emission lines diminished in 2006 \citep{Denney2014}, while reappeared with a low amplitude in 2017 \citep{Raimundo2019}. There was also a large continuum luminosity variation in optical, UV, and X-ray wavelengths. The central AGN brightened by a factor of $\sim$10 between 1970s and 1990s, then faded by a factor of $\sim$100 at optical and UV wavelengths between the mid-1990s and 2013 \citep{Denney2014}. 

At radio wavelengths, Mrk~590 has not only a radio-emitting host galaxy but also a kpc-scale compact radio nucleus \citep{Koay2016VLBA}. The Multi-Element Radio Linked Interferometer Network (MERLIN) observations on 1995 December 9 reported that the compact nucleus had a total flux density $6.0 \pm 0.2$~mJy at 1.6~GHz and a size of $\la$90~mas \citep{Thean2001}. The radio nucleus had a correlation amplitude of $\sim$5.0~mJy at 2.3~GHz on the 275~km Parkes-Tidbinbilla Interferometer baseline \citep{Roy1994}. The Karl G. Jansky Very Large Array (VLA) multi-band observations in A-array configuration made by \citet{Koay2016VLBA} reported that its radio nucleus had a flux density of $2.75 \pm 0.08$~mJy at 1.8~GHz and a very flat spectrum between 1.8 and 8.4~GHz on 2015 June 23. Combined with early sub-arcsecond-resolution radio observations \citep{Ulvestad1984, Becker1995, Kukula1995, Kinney2000}, \citet{Koay2016VLBA} showed that the radio nucleus also exhibited a coincident outburst peaking between the 1980s and 1990s. The high-resolution Very Long Baseline Array (VLBA) observations at 1.6 and 8.4~GHz found a pc-scale compact radio core \citep{Koay2016VLBA}. To search for potential faint jet activity launched by the accreting SMBH during and after the outburst, we performed deep very long baseline interferometry (VLBI) observations of Mrk~590 with the European VLBI Network (EVN). 

\begin{table*}
\caption{Short summary of the EVN observations of Mrk~590 at 1.66~GHz. The two-letter codes for the participating stations are explained in Section~\ref{sec:obs}. }
\label{tab:exp}
\begin{tabular}{cccc}
\hline
 Date       &  Participating EVN stations                    &  Project code (mode) & Phase-referencing \\
\hline
2015 Jan 13 & \texttt{EF, WB, JB, HH, ON, TR}                & EY022 ($e$-VLBI)       & Poor \\  
2015 May 13 & \texttt{EF, WB, JB, HH, ON, TR, MC, SH, T6}    & EY022B ($e$-VLBI)      & Poor \\
2015 Oct 16 & \texttt{EF, W1, JB, HH, ON, TR, MC, SH, T6, SV, BD, ZC, RO, SR, UR}                                                                                               & EY023 (disk-recording VLBI) & Success \\
\hline
\end{tabular}
\end{table*}

The Letter is organised in the following sequence. We introduce our EVN observations and data reduction in Section~\ref{sec:obs} and present our VLBI imaging results of Mrk~590 in Section~\ref{sec:results}. We interpret the observed radio structure as the central SMBH accretion-ejection activity and discuss potential implications from our findings in Section~\ref{sec:discussion}. Throughout the paper, a standard $\Lambda$CDM cosmological model with $H_{\rm 0}$~=~71~km\,s$^{-1}$\,Mpc$^{-1}$, $\Omega_{\rm m}$~=~0.27, $\Omega_{\Lambda}$~=~0.73 is adopted. The VLBI images have a scale of 0.5~pc\,mas$^{-1}$.

\section{Observations and data reduction}
\label{sec:obs}

We observed Mrk~590 three times with the EVN at 1.66~GHz in 2015. The basic experiment parameters are listed in Table~\ref{tab:exp}. The participating stations were Effelsberg (\texttt{EF}), phased-up array of the Westerbork Synthesis Radio Telescope (\texttt{WB}), Westerbork single antenna (\texttt{W1}), Jodrell Bank MK II (\texttt{JB}), Hartebeesthoek (\texttt{HH}), Onsala (\texttt{ON}), Toru\'n (\texttt{TR}), Medicina (\texttt{MC}), Shanghai (\texttt{SH}), Tianma (\texttt{T6}), Svetloe (\texttt{SV}), Badary (\texttt{BD}), Zelenchukskaya (\texttt{ZC}),  Robeldo (\texttt{RO}, 70-m dish, only left-circular polarisation available), Sardinia (\texttt{SR}) and Urumqi (\texttt{UR}). At the time of the observations, both \texttt{T6} and \texttt{SR} were newly constructed telescopes. They ran the VLBI observations in the risk-sharing mode. All the experiments used the available maximum data rate 1024~Mbps (16~MHz filters, 2~bit quantisation, 16~sub-bands in dual polarisation). The first two experiments were performed in the $e$-VLBI mode \citep{Szomoru2008}. The raw data were transferred to JIVE (Joint Institute for VLBI, ERIC) via broad-band connections and then correlated in real time by the EVN software correlator \citep[\textsc{SFXC},][]{Keimpema2015}. The last experiment EY023 was carried out in the traditional disk-recording mode to include more telescopes and gain a better ($u$, $v$) coverage. All the experiments were correlated with typical correlation parameters (1 or 2-s integration time, 32 or 64~points per subbands) used for continuum experiments.

During the observations of Mrk~590, we selected J0216$-$0118 as the phase-referencing calibrator. The calibrator is 39~arcmin away from 
our target. Its correlation position is RA~=~02$^{\rm h}$16$^{\rm m}$05$\fs$66384, Dec~=~$-$01$\degr$18$\arcmin$03$\farcs$39692 (J2000, $\sigma_{\rm ra}=0.15$~mas, $\sigma_{\rm dec}=0.30$~mas), in agreement with the positions reported by the 3rd realisation of International Celestial Reference Frame \citep[ICRF3,][]{Charlot2020} and the second data release (DR2) of the optical \textit{Gaia} astrometry \citep{Gaia2018}. It has an unresolved structure with a correlation amplitude of $\sim$0.1~Jy on the long baselines in the VLBI observations at 2.3~GHz \citep{Pushkarev2012}. The cycle time of the nodding observations was $\sim$4.5~min ($\sim$0.5~min for J0216$-$0118, $\sim$3~min for Mrk~590, $\sim$1~min for two gaps). We also observed bright calibrators (B0234$+$285, B0528$+$134, and 3C~84) as the fringe finders and the bandpass calibrators. To verify the phase-referencing calibration quality in the last epoch, we
alternatively observed a pair of nearby calibrators J0217$-$0121 and J0216$-$0105, which have a quite similar angular separation (30~arcmin).

The correlated data were calibrated with the National Radio Astronomy Observatory (NRAO) software package Astronomical Image Processing System \citep[\textsc{aips},][]{Greisen2003}. (1) We excluded one eighth of channels on each side because of their very low correlation amplitude when the data were loaded into \textsc{aips}. (2) Because of removing these side channels, we re-normalised the cross-correlation amplitude of the visibility data with the \textsc{aips} task \texttt{ACCOR} according to the auto-correlation amplitude. (3) The amplitude calibration was initially done with properly smoothed antenna monitoring data (system temperatures and gain curves) or nominal system equivalent flux densities in case of missing these data. We also corrected some sub-band dependent constant amplitude errors derived by the amplitude self-calibration in \textsc{difmap} \citep{Shepherd1994} in the later re-run of the calibration with the \textsc{aips} task \texttt{SNCOR}. (4) The ionospheric dispersive delays were corrected according to maps of total electron content provided by Global Positioning System (GPS) satellite observations. (5) The phase errors due to the antenna parallactic angle variations were removed. (6) We aligned the phases across the subbands via iteratively running fringe-fitting with a short scan of the calibrator data. After the phase alignment, we combined all the subbands in the Stokes $RR$ and $LL$, ran the fringe-fitting with a sensitive station as the reference station (\texttt{EF} or \texttt{WB}) and applied the solutions to all the related sources. (7) The bandpass calibration was performed. All the above calibration steps were scripted in the \textsc{ParselTongue} interface \citep{Kettenis2006}. 

The de-convolution was performed in \textsc{difmap} \citep{Shepherd1994}. The calibrator imaging procedure was performed through a number of iterations of model fitting with a group of delta functions, i.e., point source models, and the self-calibration in \textsc{difmap}. We re-ran the fringe-fitting and the amplitude and phase self-calibration in \textsc{aips} with the input source model made in \textsc{difmap}. All these solutions were also transferred to the target data via linear interpolation. 

The phase-referencing calibrator J0216$-$0118 shows a compact jet structure quite close to a point source. It had total flux densities $\sim$99, $\sim$108, and $\sim$115~mJy at 1.66~GHz in the three epochs. The flux densities were slightly lower than the value (128~mJy) reported by ICRF3 observations at 2.3~GHz on 2015 January 23.

Our target source Mrk~590 was detected in all the epochs. However, we noticed that there were significant phase fluctuations on rather short time scales $\ga$5~min in the first two epochs because of the unexpected severe space weather near the peak of the 11-yr solar activity cycle and the not-too-large angular distances (98$\degr$ and 24$\degr$) to the Sun. These phase issues caused poor phase connections between the calibrator and the target and significant phase coherence loss ($\ga$40 per cent). The dirty maps had rather low peak brightness: $\sim$0.36 and $\sim$0.35~mJy\,beam$^{-1}$ with uniform grid weighting. Because of the non-optimal ($u$, $v$) coverage, the dirty beam had multiple strong side-lobes. In view of the two problems, it was hard to detect any secondary feature (e.g. the component N in Fig.~\ref{fig:mrk590}). Thus, the results were not reported in the paper. In the last epoch, \texttt{SR} and \texttt{UR} had no fringes, \texttt{SH} had a linear polarisation signal, \texttt{T6} had significant amplitude fluctuations on rather short time scale of a few seconds in the parallel-hand correlation data of Stokes $LL$, \texttt{TR} had a problem with tracking Mrk~590 in the last three hours and \texttt{ON} had significantly residual errors likely due to some unexpected ionospheric activity at the high latitude. All these problematic data were excluded. Moreover, we noticed that \texttt{T6} had a certain systematic noise in the remaining Stokes $RR$ data and lowered down its data weight by a factor of 0.25 with the \textsc{aips} task \texttt{WTMOD}. Because most stations ran the observations (19h -- 3h UT) at night (the Sun distance 164$\degr$), the phase varied slowly and smoothly, and the phase interpolations worked precisely in the last epoch. The phase-referencing calibration between the calibrators J0217$-$0121 and J0216$-$0105 also worked as expected.         

\section{EVN imaging results on 2015 October 16}
\label{sec:results}

\begin{figure}
\includegraphics[width=\columnwidth]{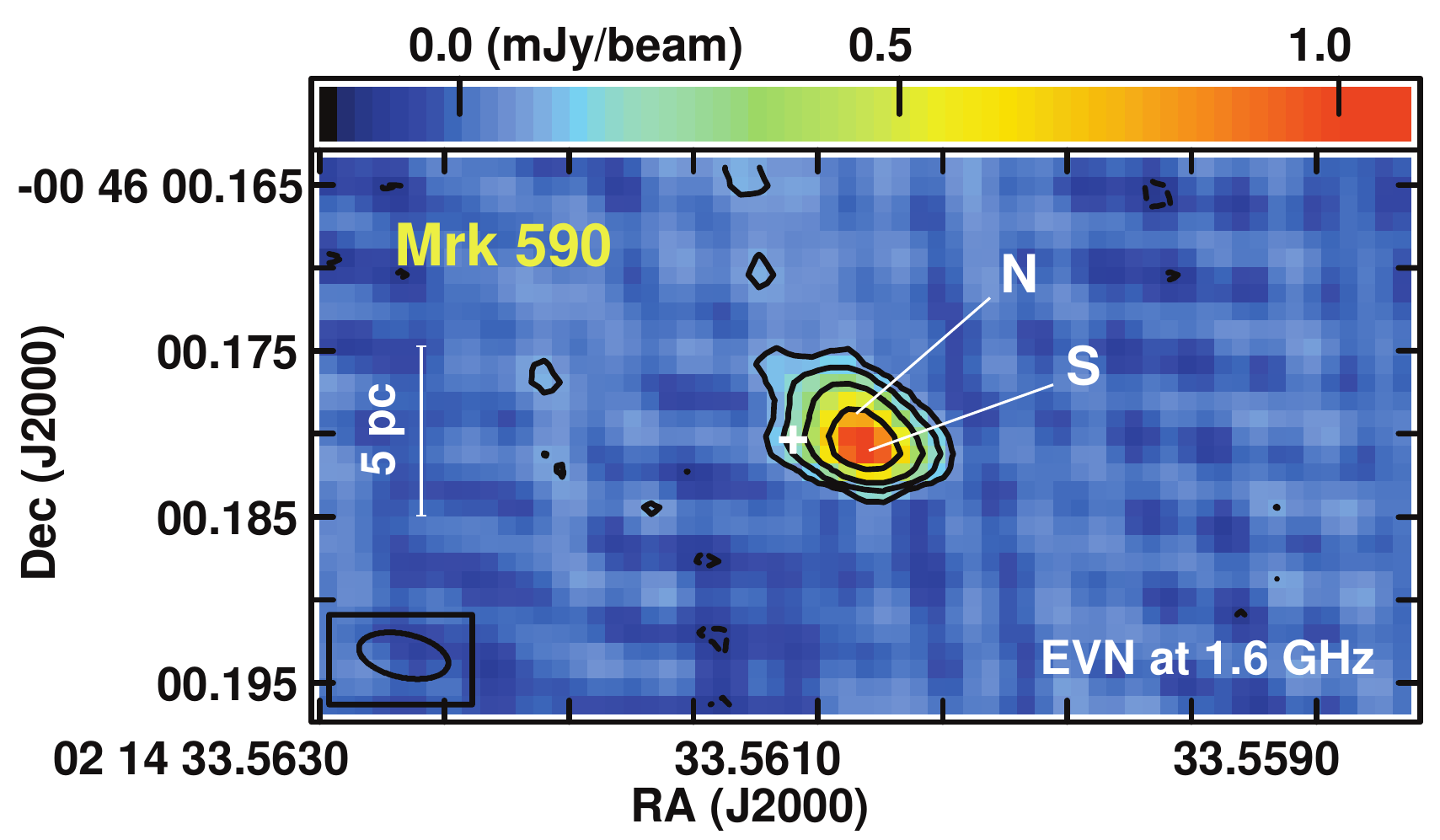}  \\
\caption{
The EVN 1.6-GHz image of the nearby changing-look Seyfert galaxy Mrk~590 on 2015 October 16. The contours start from 2.5$\sigma$ and increase by factors of $-$1, 1, 2, 4 and 8.  The map was made with uniform weighting. The beam full width at half maximum (FWHM) is 5.48~$\times$~2.62~mas$^{2}$ at position angle (PA) $78\fdg0$. The first contour is 0.083~mJy\,beam$^{-1}$ and the peak brightness is 0.107~mJy\,beam$^{-1}$. The white cross denotes the optical centroid reported by the \textit{Gaia} astrometry. }
\label{fig:mrk590}
\end{figure}

Fig.~\ref{fig:mrk590} shows the \textsc{clean} map of Mrk~590 made with the uniform grid weighting and without running the phase self-calibration. Enabled by the high resolution in particular in the north-south direction, Mrk 590 shows a significantly elongated structure. To quantify the radial extension, we decomposed the structure into two components, marked as N and S. After peeling off the peak feature S, the extension N has a signal to noise ratio (SNR) of about seven in the residual map. During the de-convolution process, we added windows carefully and timely to avoid cleaning components at the positions of strong (80 percent) side lobes. The \textsc{clean} algorithm gives a total flux density of $1.64 \pm 0.16$~mJy. In the flux density uncertainty, we included ten per cent of the flux densities as the systematic errors. Compared to the value (2.75~$\pm$~0.07~mJy at 1.8~GHz) observed by the A-configuration of the Jansky VLA on 2015 June 23 \citep{Koay2016VLBA}, the EVN image recovers about 60 per cent of its total flux density. To search for diffuse components, we also tried naturally weighting and added different Gaussian tappers to weight down long-baseline data. While, there is only a hint (SNR $<$5) of a faint ($\la$0.4~mJy) and diffuse component extending further toward north from the component N. The entire VLBI structure fully agrees with the early 1.6-GHz deep (sensitivity 0.1~mJy\,beam$^{-1}$) MERLIN constraint on the size $\la$90~mas \citep[][]{Thean2001}. In the 1.6-GHz VLBA image \citep{Koay2016VLBA}, both the total flux density and the peak brightness were likely over-estimated by a factor of about two because the phase self-calibration was tried to remove significant residual phase errors. 

We fit two point sources to the visibility data. The best-fit point sources have flux densities $1.17 \pm 0.12$~mJy for S and $0.63 \pm 0.06$~mJy for N, and a separation of $2.8 \pm 0.3$~mas (projected distance $\sim$1.4~pc) at PA = $13.0 \pm 6$~deg. We also fit the data to a Circular Gaussian model. The derived size is $2.1 \pm 0.1$~mas. The centroid position of the radio structure is reported in Table~\ref{tab:pos}. Moreover, the optical \textit{Gaia} DR2 astrometry results with a point source model are also listed in Table~\ref{tab:pos}. The EVN astrometry results are consistent with the optical Gaia DR2 position \citep{Gaia2018} and the VLBA astrometry results at 1.6 and 8.4~GHz \citep{Koay2016VLBA}. Because of instrumental limitations, it is hard to fit the structure to the more complex model, e.g. two circular Gaussian models or an elliptical Gaussian model. 

The average brightness temperature $T_{\rm b}$ of the entire radio structure is estimated \citep[e.g.,][]{Condon1982} as
\begin{equation}
T_\mathrm{b} = 1.22\times10^{9}\frac{S_\mathrm{int}}{\nu_\mathrm{obs}^2\theta_\mathrm{size}^2}(1+z),
\label{eq1}
\end{equation}
where $S_\mathrm{int}$ is the integrated flux density in mJy, $\nu_\mathrm{obs}$ is the observing frequency in GHz, $\theta_\mathrm{size}$ is the FWHM of the best-fit circular Gaussian model in mas, and $z$ is the redshift. Because the source is unresolved in the east-west direction, the size estimate is very likely an upper limit.  So, the brightness temperature estimate should be taken as a lower limit, i.e. $T_{\rm b} \ga 1 \times 10^8$~K.

\begin{table*}
\caption{List of the high-precision astrometry results of Mrk~590 presented by the optical \textit{Gaia} DR2 \citep{Gaia2018} and our EVN phase-referencing observations at 1.66~GHz. The position errors $\sigma_{\rm ra}$ and $\sigma_{\rm dec}$ include both the formal error $\sigma_{\rm f}$ and the systematic error $\sigma_{\rm s}$. }
\label{tab:pos}
\begin{tabular}{cccccc}
\hline
Technique                                & Right Ascension & $\sigma_{\rm ra}$ ($\sigma_{\rm f}$, $\sigma_{\rm s}$) & Declination &  $\sigma_{\rm dec}$ ($\sigma_{\rm f}$, $\sigma_{\rm s}$) & Comment on $\sigma_{\rm s}$          \\
                                         &  (J2000)        &  (mas)  &      (J2000)    &   (mas)  & \\  
\hline
Optical \textit{Gaia} astrometry (DR2)   & 02$^{\rm h}$14$^{\rm m}$33$\fs$561095   & $\pm0.7\pm4.0$  & $-$00$\degr$46$\arcmin$00$\farcs$18013   & $\pm0.7\pm4.0$  & Extended nucleus structure   \\
VLBI differential astrometry             & 02$^{\rm h}$14$^{\rm m}$33$\fs$560827   & $\pm0.2\pm1.0$  & $-$00$\degr$46$\arcmin$00$\farcs$18064   & $\pm0.3\pm1.0$  & Core shift of ICRF3 J0216$-$0118  \\
\hline
\end{tabular}
\end{table*}

\section{Discussion}
\label{sec:discussion}
\subsection{A low-radio-power jet}
\label{sec:jet}
The elongated radio structure can be naturally interpreted as a faint jet in Mrk~590. According to the positional consistency between our VLBI astrometry and the \textit{Gaia} astrometry, the structure represents the parsec- and sub-parsec-scale AGN activity powered by the central accreting SMBH. Based on the total flux densities: $3.0 \pm 0.2$~mJy at 8.4~GHz observed by \citet{Koay2016VLBA} with the VLBA, and  $1.7 \pm 0.2$~mJy at 1.6~GHz by us with the EVN, it has a slightly inverted radio spectrum of $S_\nu \propto \nu^{0.35 \pm 0.08}$ and thus hosts a partially synchrotron self-absorbed jet base.

The jet direction is fully consistent with the direction of the nuclear gas outflows extending up to $\sim$1.5~kpc in the north-south direction \citep{Raimundo2019, Schmitt2003}.  The outflows are very likely launched from the underlying accretion flow, as predicted by the SMBH accretion theory \citep[e.g.][]{Blandford1999, Yuan2014}, and are expected to escape mainly along the jet direction \citep[e.g., numerical simulations by][]{Yuan2012, Yuan2015}. Together with the Doppler shifts of the emission lines of the outflows \citep{Raimundo2019}, the components N and S can be identified as the approaching jet and the jet base respectively. Faint radio cores are also detected in many Seyfert galaxies \citep[e.g.][]{Giroletti2009, Bontempi2012, Gabanyi2018} and quasars \citep[e.g.][]{Yang2012}. The faint jet in Mrk~590 has a very low radio luminosity of $\sim4\times10^{37}$~erg\,s$^{-1}$.  Thus, it can be identified as a low-radio-power jet \citep[e.g.][]{Kunert2010, An2012}. According to the fundamental plane relation of black hole activity \citep{Kording2006}, its low radio luminosity can be reasonably explained as a consequence of its relatively low black hole mass and low X-ray luminosity \citep{Koay2016VLBA} in the low accretion rate state. Compared to the radio luminosities of nearby galaxies, e.g. $\sim$10$^{34}$--10$^{40}$ erg~s$^{-1}$ \citep[280 galaxies,][]{Baldi2021}, the radio luminosity of Mrk 590 is a typical value.

The linear structure has a total radio luminosity below the maximum luminosity, $L_{\rm R}$ $\sim$10$^{38.7}$~erg\,s$^{-1}$, observed in the young supernovae \citep{Weiler2002}. However, it cannot be explained as young supernovae or supernova remnants \citep[e.g.][]{Varenius2019}. Optical observations show that there is no sign of star-forming activity in the nuclear region \citep{Raimundo2019}. Furthermore, the molecular gas mass in the inner 150~pc is very low, $\la 1.6 \times 10^5$ M$_{\sun}$ \citep{Koay2016ALMA}.

\subsection{Implications from the jet activity}
\label{sec:outburst}
Mrk~590 underwent a giant outburst and then a slow fading from radio to X-ray bands over the last fifty years \citep{Denney2014, Koay2016VLBA}. Among the known changing-look AGN, Mrk~590 is the first case displaying the coincident radio variability \citep{Koay2016VLBA}. At radio, it reached $6.3 \pm 0.3$~mJy at 1.4~GHz with the sub-arcsec-resolution observations in 1995 \citep{Thean2001} and faded to $3.4 \pm 0.1$~mJy at 1.3~GHz with the arcsec-resolution observations in 2015 \citep{Koay2016VLBA}. The findings of the elongated jet provides strong evidence of significant accretion-ejection activity in Mrk~590. It is not clear whether the entire jet structure resulted from the outburst. The stable jet base S might be formed before the outburst. If the long-term rising and fading radio activity is associated with the jet component N, it would have a life of $\la45$~yr, show an apparent separation speed of $\ga$0.1$c$ with respect to the component S, and keep fading slowly for the rest of its life. Since the intensive accretion was a very short active phase, the jet might be short of the energy supply and then die rapidly as a short-lived jet \citep{Woowska2017}. To strengthen or exclude the association, it requires future multi-epoch deep VLBI observations to search for its proper motion and flux density variability.

Mrk~590 had a low accretion rate of $L_{\rm bol}/L_{\rm Edd} \sim 10^{-3}$ in the low luminosity state \citep{Koay2016VLBA} while reached a very high accretion rate of $L_{\rm bol}/L_{\rm Edd} \sim 10^{-1}$ \citep{Peterson2004} during the outburst. Based on a small sample of faded changing-look quasars, \citep{Ruan2019} has recently found a V-shaped evolution pattern in the plot of the UV-to-X-ray spectral index versus the Eddington ratio. The behaviour is in agreement with the prediction of the AGN accretion state transition generated by \citet{Sobolewska2011} based on stellar-mass black holes in X-ray binaries. The critical accretion rate to discriminate high and low states is likely $\sim$10$^{-2}$ \citep{Ruan2019}. This is also consistent with the value observed in stellar-mass black hole X-ray binaries \citep[e.g.][]{McClintock2006}. If the unified X-ray outburst model in black hole X-ray binaries \citep{Fender2009} is applicable to Mrk~590, the early brightening would represent a transition from the low to high accretion rates, and the component N might be ejected during the state transition. If the ejection activity resembles the situation observed by \citet{Marscher2002} in 3C~120, multiple ejection events might have occurred not only during the outburst but also during the follow-up fading stage.

The rapid disappearance and re-appearance of broad Balmer lines in Mrk~590 most likely results from the variable accretion instead of line-of-sight obscuration by gas and dust \citep{Denney2014, Koay2016VLBA, Mathur2018}. This is mainly because of the detection of the coincident radio variability \citep{Koay2016VLBA} and the abscence of intrinsic absorption in the X-ray spectrum \citep[e.g.][]{Mathur2018}. Our finding of the small faint jet provides additional support for the variable accretion scenario.


\section*{Acknowledgements}
$e$-VLBI research infrastructure in Europe is supported by the European Union’s Seventh Framework Programme (FP7/2007-2013) under grant agreement number RI-261525 NEXPReS.
The European VLBI Network (EVN) is a joint facility of independent European, African, Asian, and North American radio astronomy institutes. Scientific results from data presented in this publication are derived from the following EVN project codes: EY022 and EY023. 
The research leading to these results has received funding from the European Commission Seventh Framework Programme (FP/2007-2013) under grant agreement No. 283393 (RadioNet3).
This work has made use of data from the European Space Agency (ESA) mission {\it Gaia} (\url{https://www.cosmos.esa.int/gaia}), processed by the {\it Gaia} Data Processing and Analysis Consortium (DPAC, \url{https://www.cosmos.esa.int/web/gaia/dpac/consortium}). Funding for the DPAC has been provided by national institutions, in particular the institutions participating in the {\it Gaia} Multilateral Agreement.
This research has made use of the NASA/IPAC Extragalactic Database (NED), which is operated by the Jet Propulsion Laboratory, California Institute of Technology, under contract with the National Aeronautics and Space Administration.
This research has made use of NASA’s Astrophysics Data System Bibliographic Services.

\section*{Data Availability}
The correlation data of EY022 and EY023 underlying this article are available in the EVN data archive (\url{http://www.jive.eu/select-experiment}). The calibrated visibility data will be shared on reasonable request to the corresponding author.
 



\bibliographystyle{mnras}
\bibliography{Mrk590} 








\bsp	
\label{lastpage}
\end{document}